\documentclass{aa}  

\usepackage{graphicx}
\usepackage{txfonts}
\begin{document}

\title{Periodicities in an active region correlated with Type III radio bursts
observed by Parker Solar Probe}

\titlerunning{Periodicities in Type III  and euv}

\author{Cynthia Cattell\inst{1}
          \and {Lindsay Glesener}\inst{1}
          \and {Benjamin Leiran}\inst{1}
          \and{Keith Goetz}\inst{1}
          \and {Juan Carlos Mart\'{i}nez Oliveros}\inst{2}
          \and{Samuel T. Badman}\inst{2,3}
          \and {Marc Pulupa}\inst{2}
          \and {Stuart D. Bale} \inst{2,3}
          }

\institute{School of Physics and Astronomy, University of Minnesota, 116 Church St. SE  Minneapolis, MN 55455 USA\\
              \email{cattell@umn.edu}
         \and
            {Space Sciences Laboratory, University of California, Berkeley, Berkeley,CA 94709 USA}
        \and
        (Department of Physics, University of California, Berkeley, Berkeley, CA 94709}

 \date{Received ; accepted }

 
  \abstract
   {Periodicities have frequently been reported across many
wavelengths in the solar corona. Correlated periods of
$\sim$5 minutes, comparable to solar p-modes, are suggestive
of coupling between the photosphere and the corona.}
  {Our study investigates whether there are
correlations in the periodic behavior of Type III radio bursts,
indicative of non-thermal electron acceleration processes, and coronal
EUV emission, assessing heating and cooling, in an active region when there
are no large flares.}
  {We use coordinated observations of Type III radio bursts from
the FIELDS instrument on Parker Solar Probe (PSP), of extreme
ultraviolet emissions by the Solar Dynamics Observatory (SDO)/AIA and
white light observations by SDO/HMI, and of solar flare x-rays by
Nuclear Spectroscopic Telescope Array (NuSTAR) on April 12, 2019.
Several methods for assessing periodicities are utilized and compared to
validate periods obtained.}
  {Periodicities of $\sim$ 5 minutes in the EUV in
several areas of an active region are well correlated with the
repetition rate of the Type III radio bursts observed on both PSP and
Wind. Detrended 211A and 171A light curves show periodic profiles in
multiple locations, with 171A peaks lagging those seen in 211A. This is
suggestive of impulsive events that result in heating and then cooling in the lower corona. NuSTAR x-rays
provide evidence for at least one microflare during the interval of Type
III bursts, but there is not a one-to-one correspondence between the
x-rays and the Type-III bursts. Our study provides evidence for periodic
acceleration of non-thermal electrons (required to generate Type III
radio bursts) when there were no observable flares either in the x-ray
data or the EUV. The acceleration process, therefore, must be associated
with small impulsive events, perhaps nanoflares.}
   {}


\maketitle

\section{Introduction}
{\label{introduction}}

Quasi-periodic variations with periods ranging from seconds to tens of
minutes have long been reported for many phenomena in the active and
quiescent solar corona, starting from the first detection of correlated
periodicities in solar flare x-rays and microwaves \citep{parks1969}.
Examples include rapid variations in Type III
radio bursts (Mangeney and Pick, 1989; Ramesh et al., 2005). In the
extreme ultraviolet (EUV), the extensive observations of these
periodicities, often interpreted as the signature of magnetohydrodynamic
(MHD) waves, have led to the development of coronal seismology, to
assess properties of the corona \citep{nakariakov2005, kupriyanova2020, roberts1984, roberts2000, demoortel2012}. Long-lived 3 to 5 minute pulsations are also
observed in sunspots \citep{sych2012, battams2019, sych2020}. 
Quasi-periodic variations in x-rays and gamma-rays from
solar and stellar flares are also observed \citep{vandoorsselaere2016, kupriyanova2020, inglis2015, dennis2017, hayes2020}.

Several studies have described correlated periodicities in various
combinations of Type III radio bursts, hard x-rays, EUV emissions,
microwaves, and sunspots at periods of minutes. Type III bursts are of
particular interest because they provide information on acceleration of
non-thermal electron beams. \citet{innes2011} reported correlations
between~ $\sim$3 minute periodicities in Type III radio
bursts and coronal jets observed in 211 \AA, which were possibly related
to 3-minute oscillations in sunspot brightness. Most other studies have
described periodicities in association with large flares. Oscillations
in Type III radio waves, hard x-rays, and jets at $\sim$4
minute periods were reported by \citet{li2015}.  \citet{kumar2016} 
found $\sim$3 minute pulsations in hard x-rays, microwave
emission, Type III bursts and a nearby sunspot.

 \citet{foullon2010} 
found periods of $\sim$10 minutes in Type IIIs and x-rays,
with longer $\sim$18 minute periods in the few GHz radio
emissions (interpreted as due to non-thermal gyroresonance). Shorter
periods ($\sim$100s) were reported by \citet{kumar2017} in
correlations between coronal fast mode waves, Type III and IV radio bursts,
microwaves, and thermal x-rays. \citet{kupriyanova2016} described 
$\sim$40 s variations in hard x-rays, microwaves and Type III
waves, consistent with modulation of non-thermal electron acceleration,
but not thermal processes.

Explanations for the periodicities include MHD waves (Alfven waves and
fast or slow mode magnetosonic waves), modulation of reconnection at
flare sites via intrinsic processes, current sheet structure \citep{demoortel2012, mclaughlin2018, aschwanden2005}, or coupling of solar p-modes to coronal waves \citep{zhao2016}. For events that include non-thermal x-rays and/or Type III radio
bursts, periodicities are most often attributed to modulated
reconnection. Theoretical and modeling studies of reconnection with
periodicities, which can be due to intrinsic loading/unloading
timescales, modulation by MHD waves, or other processes,~ include \citet{murray2009, mclaughlin2012,  thurgood2017}.  An
alternate approach attributes quasi-periodic behavior to stochastic
processes \citep{veronig2000, aschwanden2016, eastwood2010}.

In this report, we describe observations of repetitive Type III bursts
observed by Parker Solar Probe (PSP) on April 12, 2019, and their
correlation with periodic rapid heating and cooling in the 211 \AA\ and 171 \AA\
bandpass filters ($\sim$2 MK and $\sim$0.6 MK peak
temperature responses, respectively) on the Solar Dynamics Observatory
(SDO) / Atmospheric Imaging Assembly (AIA). Section 2 describes the data
sets and analysis techniques; Section 3 shows the observations; and
Section 4 discusses the results, comparisons to other studies and
interpretation in terms of possible physical models.

\begin{figure*}[h!]
\begin{center}
    
\includegraphics[width=.95\textwidth]{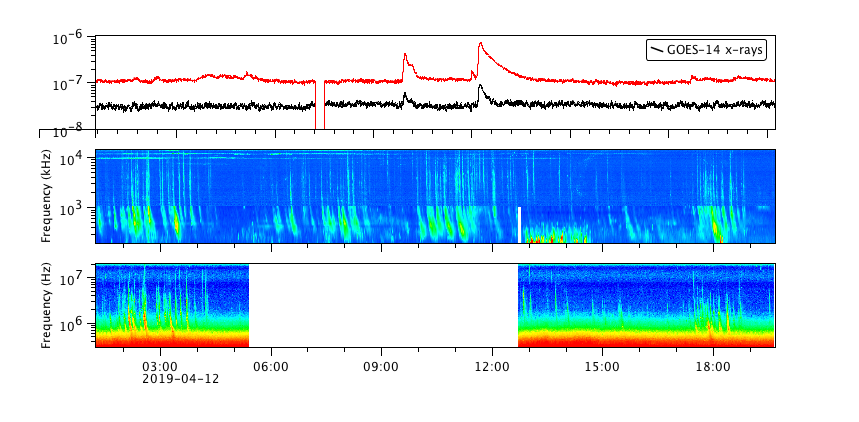}
\caption{Overview of (top) x-rays observed by GOES-14, and Type III radio
bursts (300 kHz to 20 MHz) observed by~ (middle)~ Wind~ and~ (bottom) Parker Solar
Probe for April 12, 2019 0100 to 2000 UT.  Times are those at which the emission was recorded on the respective spacecraft; the PSP times have not been shifted to 1 AU.
\label{time_profile_long}
}
\end{center}
\end{figure*}

\section{Data sets}
{\label{data}}

We focus on an interval on April 12, 2019, when simultaneous data were
obtained by NuSTAR, PSP, Wind, and SDO. Radio data were obtained with the Radio Frequency Spectrometer (RFS)\citep{pulupa2017}, part of the PSP FIELDS suite \citep{bale2016}. This interval is at the end of the
second encounter, so the sample rate was low, $\sim$1 sample
per 55 s. We also examined higher rate (1 sample per 7 s) data for
intervals with similar periodic bursts. We also utilize radio data from
Wind/Waves \citep{bougeret1995}, from RAD1 (20-1040 kHz) and RAD2
(1.075-13.825 MHz) at 1 sample per 16 s.

Extreme ultraviolet data from SDO/AIA
\citep{lemen2012} and magnetic field information from the
Helioseismic and Magnetic Imager \citep[HMI;][]{schou2012} are utilized
to examine periodicities and solar structures. AIA data were obtained
in 7 wavelengths at a 12 s cadence over the full sun. Analysis of AIA
data is focused on five areas within active region 12738 (NOAA
designation).

During the interval of interest, based on field
line tracing using a combination of the Potential Field Source Surface
model and the Parker spiral (see \citet{badman2020} for a description of the method), PSP
and Wind both map closer to the smaller active region near 60 degrees longitude,
which was not observable by SDO at this time. \citet{harra2020} study this smaller active region (AR 12737) and conclude from the dynamics that it may be contributing to the population of Type III bursts observed, although they study an earlier interval from March 31-April 6th, during which time AR 12738 was behind the limb as viewed from Earth.  PFSS mapping also suggests that there was no direct magnetic connection to either active region and thus no in situ observations of electron beams are expected (or indeed observed) at PSP or Wind. One study \citep{krupar2020} used radio triangulation with STEREO and Wind to show that in at least one case study of a large type III burst, the electron beam appears to be consistent with a Parker spiral emerging from AR 12378. \citet{Pulupa_2020}, who examined the full encounter, also associated the bursts with AR 12738, citing the consistency of the orientation of its bipole on the solar disk with the polarization of radio emission measured in situ (although this is also true of AR 12737 since it is in the same solar hemisphere in the same solar rotation). The observations of  \citet{krupar2020} and \citet{Pulupa_2020} suggest that our comparison of the Type III bursts to periodicities in AR 12738 is appropriate. 

On April 12 and April 13, 2019, at the end of the second PSP periapsis
pass, NuSTAR observed the sun, measuring hard X-rays at energies from
2.5 to 10 keV for six subintervals when PSP obtained radio data. Several
GOES A7 to A9 class flares were seen in each of the two intervals on
April 12 for which there are PSP radio data, and two B class flares on April 13 with
PSP radio data.~See Figures \ref{time_profile_long}, \ref{time_profile}, and \ref{images} for summaries of the observations.

\section{Observations}
{\label{observations}}

An overview of the event is shown in Figure \ref{time_profile_long}, including the x-ray data
from GOES-14 (panel a),~ the Type III radio bursts seen on Wind (panel
b) and~PSP/FIELDS (panel c). During the interval from ~540 UT to 1230 UT, all PSP instruments were turned off to enable high rate science data downlink.  The quasi-periodic repetition of the Type III radio
bursts seen by both PSP and Wind is clear, with periods of
$\sim$4 to 5 minutes. Note that similar bursts are observed
intermittently for $\sim$10 days from either or both of PSP
and Wind. The interval displays some characteristics of a Type III radio
storm  \citep{fainberg1970, bougeret1984, morioka2007}.
The individual bursts are lower power and have a more limited
frequency range than flare-associated Type IIIs. The repetition rate is
longer than the range found for storms ($\sim$10s~ to 1
minute); however, this may be due to the low level of solar activity. \citet{Pulupa_2020} characterized the properties of the Type III bursts for two intervals, one before our observations (April 3 08 UT to April 4 08 UT), and one after (April 17 17:00 UT to April 18 05:00 UT), and concluded the Type IIIs were associated with Type III storms.

\begin{figure}[h!]
\begin{center}
\includegraphics[width=0.85\columnwidth]{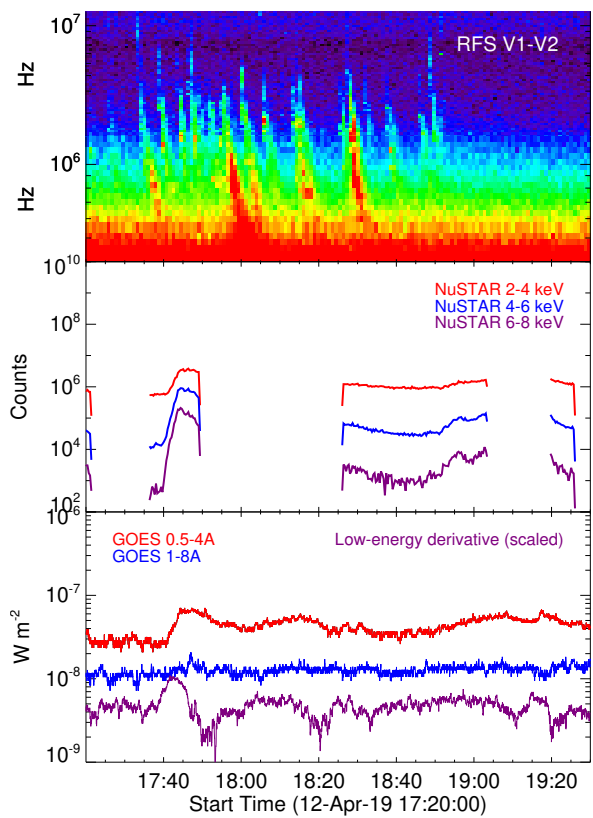}
\caption{Interval of interest from 17:20 to 19:30 UT. The top panel shows
PSP/FIELDS radio data (V12), with times propagated to 1 AU. The middle panel shows the NuSTAR x-ray data, and the bottom panel shows the GOES x-ray data.
\label{time_profile}
}
\end{center}
\end{figure}

To examine possible correlations between the low corona (as probed by
AIA) and the Type III radio bursts, we focus on the interval between
1715 and 18:45 UT. Figure \ref{time_profile} plots the PSP radio data, the NuSTAR x-rays, and the GOES x-rays together, with PSP times shifted to 1 AU. 
Although the flares do occur during times of intense radio activity and
the X-ray bursts may be related to the radio bursts, there is not a
one-to-one correspondence between the flare x-ray emissions(c and d) and
the Type III bursts.

\begin{figure*}[h!]
\begin{center}
\includegraphics[width=0.75\textwidth]{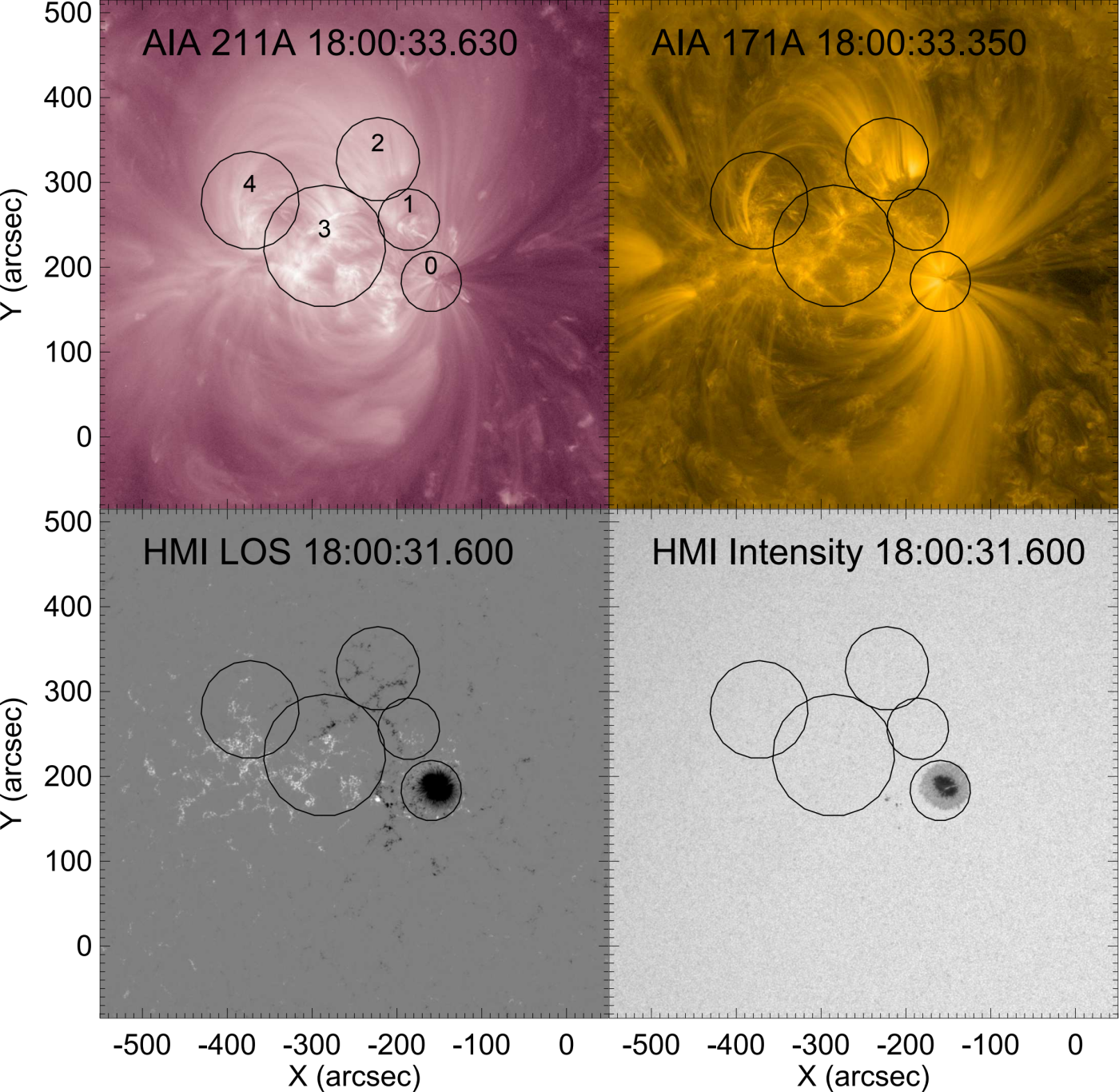}
\caption{{Regions analyzed are shown overlaid on SDO images. The top row includes two AIA filters that have sensitivity to quiescent coronal temperatures, and the bottom row includes the HMI line-of-sight magnetogram and intensity map. Images shown are at a time near the middle of the analyzed interval. At the latitude of this active region, solar rotation causes sources to drift westward at a rate of about 10 arcsec per hour.
{\label{images}}%
}}
\end{center}
\end{figure*}

\begin{figure*}[h!]
\begin{center}
\includegraphics[width=1.0\textwidth]{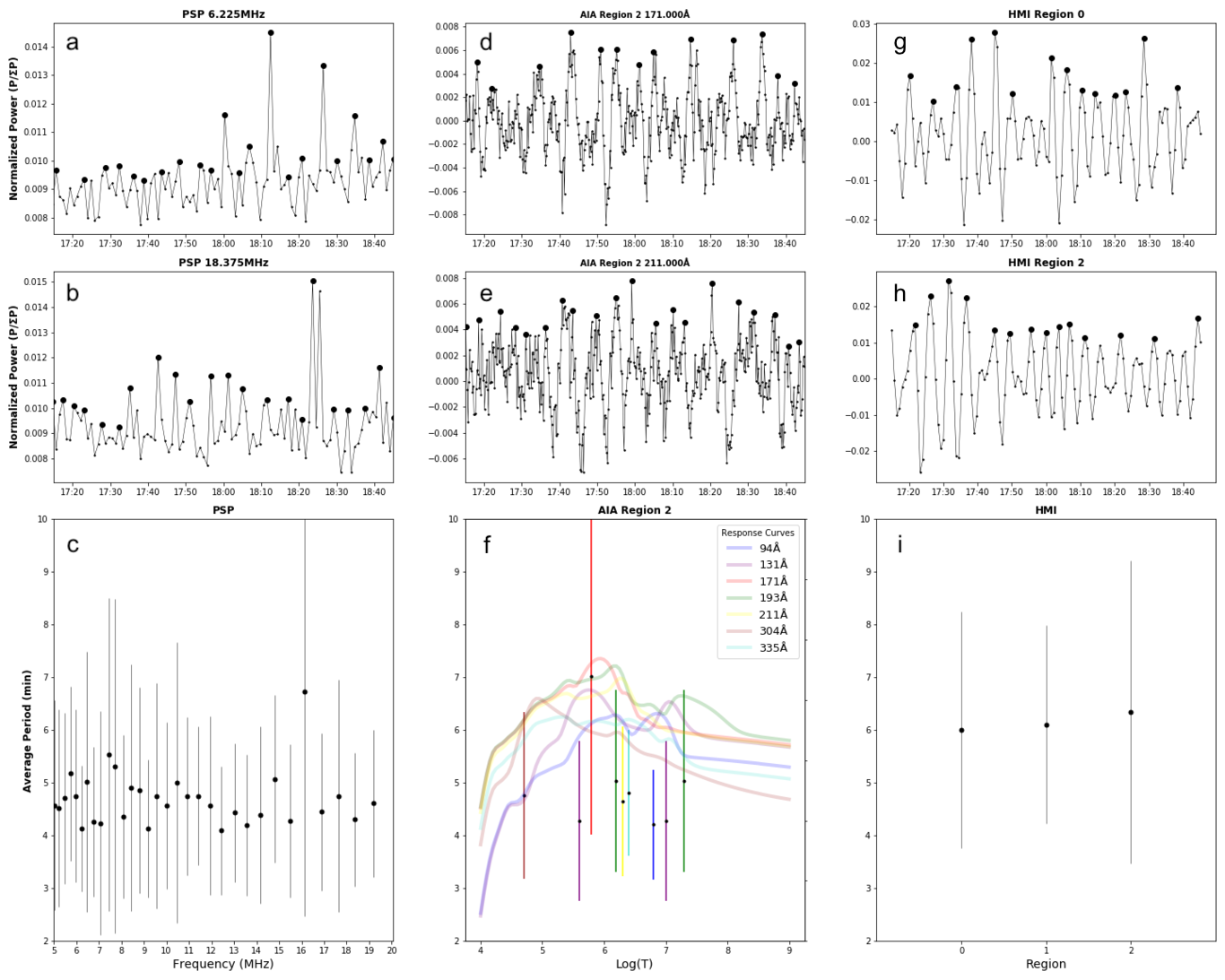}
\caption{Periodicities in PSP/FIELDS HFR radio power, AIA emission in
Region 2 (as defined in \ref{images}), and HMI intensity. (a)~ 6.225 MHz and (b) 18.375 MHz interpolated normalized~ power with black dots
indicating the identified peaks; (c) periods identified for all
frequencies \textgreater{}5 MHz; (d) AIA 171 \AA\ and (e) 211 \AA\ detrended
normalized intensity with black dots indicating the identified peaks;
and (f)~ periods identified for all AIA passbands in Region 2, versus
wavelength with temperature response curves overplotted; (g) and (h) HMI emission in two regions; and (i) periods identified in HMI data.  For panels (d) and (e), the AIA lightcurves have been detrended.  The units are residual data numbers (DNs) after a 10-minute average curve was subtracted.  In panel (f), the temperatures of the data points are those listed in Table 1 of \citep{lemen2012} for each passband (these are approximately the peaks of the temperature responses), but the colored lines show the temperature responses themselves, in arbitrary units, to give a better representation of possible temperatures for each data point.  When more than one temperature is listed in Table 1 of \citep{lemen2012}, we have included a data point for both, since we cannot distinguish them.
\label{results}
}
\end{center}
\end{figure*}

SDO/AIA measures EUV emission from the Sun in passbands defined by ten
filters, six of which are sensitive to coronal temperatures (Lemen et
al. 2012).~ Images are full-Sun at 1.2 arcsec resolution and a 12 s
cadence for each filter.~ Several diverse regions of active region 12738
were selected for individual analysis, including the region at the major
sunspot, regions where small transients were visible by eye, and some
quiet regions.~ Regions are shown in Figure \ref{images}. Within each of these
regions, AIA emission in individual filters was totaled over the region
and plotted as a function of time.~ These included all of the coronal
filters as well as the 304A filter, which is sensitive to the He II ions
typically found in the chromosphere and transition region.~ Solar
rotation was not removed, so the solar emission drifts across each
region at a slow rate ($\sim$10 arcsec per hour).~ Some
regions exhibited a periodic behavior to their time profiles in addition
to macroscopic, transient events. The time profiles were detrended so
that this periodic behavior was more apparent, as shown in Figure \ref{results}, panels (d) and (e).  Detrending was performed by subtracting a smoothed curve (with a
running average over 10 minutes) from each time profile.

Figure \ref{images} shows the active region with the five subregions indicated. The top
panels are images of 211 A and 171 A from AIA, and the bottom panels are
the line-of-sight (LOS) magnetogram and intensity map from HMI. For the EUV images, our study utilizes bandpass filter images, and as such examines only periodicity in the EUV bandpass brightnesses, and not in other properties such as line Doppler velocity or Doppler width.

\begin{figure}[h!]
\begin{center}
\includegraphics[width=1.0\columnwidth]{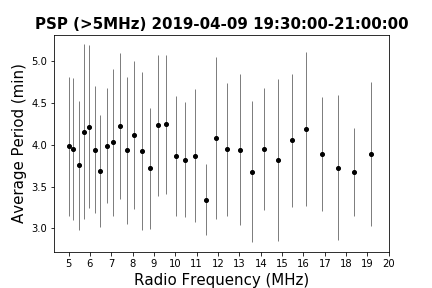}
\caption{Periodicities seen in PSP radio for earlier interval with
higher data rate.
\label{periods2}
}
\end{center}
\end{figure}

The EUV and radio data were examined for periodicities utilizing a
method that identified peaks and valleys above a threshold value in the
normalized power (PSP and Wind) or normalized detrended light curve
(SDO/AIA and HMI). The detrending time periods and intervals~for
analysis of~ AIA~ data were selected to avoid introducing artificial
periods \citep{auchere2016, dominique2018}. Periodicities
for the radio data were determined both using the measured values and
using data interpolated to the cadence of the AIA data. Figure \ref{results} shows
an example of the `peak' approach for the radio, EUV and HMI data sets.
Panels a and b show the normalized power above average for the two~
frequency bands (6.2 MHz and 18.4 MHz) from the~ PSP radio data versus
time. Panel c plots the resulting average periods versus frequency for
the HFR (\textgreater{}5 MHz) band,~ determined using a threshold of 1
for the normalized power. Panels d and e plot the detrended normalized
intensity for two AIA bands, 171 \AA\ and 211 \AA\ in Region 2; and panel
f~plots the average period versus temperature (based on the peak of the
temperature responses of each bandpass for the six coronal lines and the
one photospheric line) for Region 2, where the periodicities were most
prominent. The AIA temperature responses are overplotted; colors
identify the same lines for both periodicity and temperature response.
It should be noted that all of the AIA bandpass filters have broad
temperature responses, and some have bimodal responses. We cannot use
these data to measure a strict temperature without differential emission
measure analysis, but the peak temperature response gives a rough
estimate of which temperature range we are likely to be observing. For
the filters that have a doubly peaked response, we have plotted the
temperatures of both peaks for reference. Although some studies of
quasi-periodicities in EUV lines have identified a temperature
dependence, the periods we observe are independent of temperature to
within the error bars. Panels g and h plot the time series of the HMI
intensity for 2 regions within the EUV subregion 0, and the periods
determined for all 3 HMI regions are plotted in panel i.~~ It is clear
that the periodicity in the radio waves ($\sim$ 4-5 minutes)
is comparable to those in the AIA and HMI data.~The behavior of the EUV
emission in the other 4 regions was also examined (not shown); similar
periodicities were observed but the amplitude of the variations was
smaller. Note that periodicities in the radio, EUV and HMI data were
also determined using fast Fourier transforms (FFTs), with similar
results. We have also examined periodicities in the PSP radio data for
intervals with similar repetitive Type III bursts earlier in this pass
when higher rate data were obtained. Figure \ref{periods2} (same format as Figure \ref{results})
shows that the periodicities observed in the Type IIIs~ are very stable,
and are observed for many days. 
Repetitive Type III radio bursts observed by PSP earlier in this encounter are discussed by \citep{harra2020}.

\begin{figure}[h!]
\begin{center}
\includegraphics[width=\columnwidth]{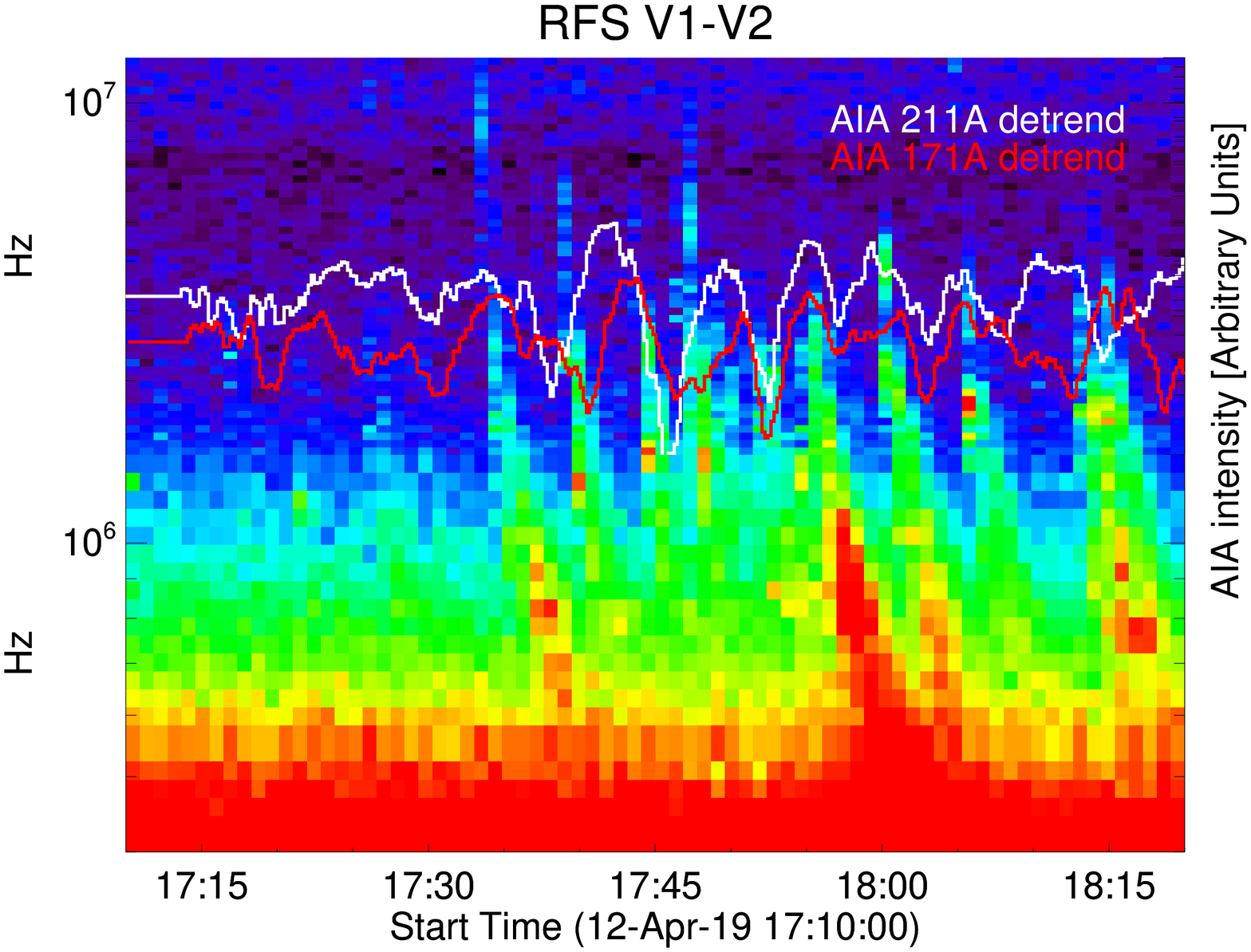}
\caption{Time series of PSP radio with detrended AIA 211 (white) and
171 (red) overplotted for Region 2.  PSP times have been propagated to 1 AU.
\label{aia_psp_overlay}
}
\end{center}
\end{figure}

The correlation between the times series of the PSP radio data and the
SDO/EUV data is shown in Figure \ref{aia_psp_overlay}, in which the 171 A (in red) and 211 A
(in white) detrended light curves are plotted on top of the PSP radio
power. The radio data have been time-shifted to account for propagation
from the solar radial position of PSP to 1 AU. For the first radio
bursts, the typical timing observed is that the radio burst is followed
by 211 A emission, with the 171 emission increasing as the 211
decreases. For most events, the rise in 211 precedes that in the 171 A.
This is suggestive of rapid heating, followed by rapid cooling. A
possible interpretation is that the heating is due to electrons
accelerated downward in the small-scale reconnection that accelerates
electrons upward to generate the radio waves.

\section{Discussion}
{\label{discussion}}

Many different mechanisms have been proposed to explain observed
periodic and quasi-periodic behavior in the EUV and radio data.~Our
study provides evidence for periodic acceleration of non-thermal
electrons (required to generate Type III radio bursts) when there were
no observable flares either in the x-ray data or the EUV. The occurrence of Type III bursts without significant flaring was also observed by \citep{harra2020} earlier in this PSP encounter.  The
acceleration process, therefore, must be associated with small impulsive
events (perhaps nanoflares) \citep{viall2011, bradshaw2012, ishikawa2017, hudson1991, klimchuk2015}, 
or with some other mechanism such 
as kinetic Alfven waves \citep{mcclements2009}. Small
acceleration events have also been seen as isolated events \citep{james2017} 
with a Type III burst and only very weak hard x-rays.~ If the
mechanism is nanoflares, the electron acceleration may involve processes
seen in flares (e.g. those described in \citet{zharkova2011}). Studies
of small microflares using NuSTAR have provided evidence for
acceleration of non-thermal electrons at energies below 7 keV \citep{glesener2020}, 
with significant collisional energy deposition that could
provide heating of the corona. Studies have predicted a range of
periodicities for nanoflares \citep{viall2011, bradshaw2012, klimchuk2015, knizhnik2020} depending on parameters
such cooling rates.

One possibility for obtaining periodicities is that MHD waves in the
corona can initiate magnetic field reconfiguration, resulting in
small-scale reconnection. It has been shown that propagating coronal
waves can destabilize active regions \citep[e.g.][]{ofman2002} 
There are many cases of fast mode waves observed in the low corona in
association with eruptions \citep{veronig2011, liu2018}, but
there has been less work on this phenomena at quiet times. Another
possibility is that waves generated by the field reconfiguration
propagate through the active region with a temperature-dependent
dispersion, for example via fast mode waves. For the fast mode (or any
other temperature-dependent mode), the observed frequencies would differ
in the AIA filters. There was no clear temperature dependence in the
periodicities for our event, suggesting that this mechanism was not
operating. Magnetic field reconnection may also occur in an inherently
periodic fashion \citep{vandoorsselaere2016, nakariakov2009}. In this case, the AIA emission represents the heating and
cooling associated with these periodic reconnection events, and all the
AIA filters, as well as the radio data, should exhibit the same
frequency, as was the case in our event.~

It is possible for the photosphere to be the ultimate source of the
periodicity in either case. \citet{demoortel2002} found different
oscillation periods for coronal loops with footpoints inside sunspots
than for ones with footpoints outside sunspots. They concluded that the
waves were not associated with flares, but rather with a driver whose
effects propagated up through the transition region. Correlations
observed by \citet{li2015} between soft x-rays and EUV jets at
$\sim$5 minute periods led them to suggest that photospheric
p-modes may lead to periodic reconnection in the corona.  \citet{depontieu2005} 
modeled magnetic flux tubes and showed that 5 minute p-mode
waves propagated into the corona, and could thus be the source of 5
minute periodicities in coronal wavelengths. The possibility that the
photosphere is the source of the periodicity we observe is consistent
with the correlation of the Type III burts and the EUV with the HMI
brightness.

A definitive determination of whether the periodicities in the EUV
observed in our event were due to MHD waves, periodic jetting or
another process would require analysis of more complex properties in the
AIA data, such as Doppler velocity \citep{depontieu2010, kupriyanova2020, demoortel2012}. \citet{liu2014} 
review identification of wave modes in the AIA data, including
fast mode waves, Kelvin-Helmholtz waves, and `mini EUV-waves.' The
latter may be relevant for our observations as they represent
small-scale, weaker waves that occur more often than larger waves.

Some studies of periodicities of minutes have concluded that modulation
of reconnection due to current sheet oscillations is most consistent
with their observations \citep{kupriyanova2016}, and that the height
over which modulated processes occur is inconsistent with MHD waves.
Other mechanisms for producing modulated reconnection have been
investigated by many researchers. \citep{nakariakov2006} discuss
enhanced reconnection associated with kinetic instabilities driven by
currents associated with fast mode waves and the interaction of magnetic
loops.  \citet{liu2011} found 3 minute fast mode waves correlated with
flare QPP, implying a causal link via wave modulation of reconnection. 
In a study of $\sim$10-20s~ QP in flares and radio waves, \citep{fleishman2008} compared properties to two models, effects of MHD
waves on radio emissions and QP injections of electrons, and concluded
that the latter better fit their observations.

Our observations are qualitatively similar to those reported by \citep{innes2011}, 
although those observations centered on very clearly
distinguishable, prominent jets.~ In the event described herein,
periodic behavior is observed most clearly in the 211 A passband (the
same as in the Innes study) but jet behavior is less prominent.~
There are certainly many jets occurring in the active region, but the
EUV oscillations are not limited to one jet-producing region. Similar
periodicities occur throughout the active region. \citet{li2015} 
described recurrent jets associated with magnetic flux cancelation and
periodic, 5 and 13 minutes, brightenings in the EUV at the jet base.
They concluded that their observations were consistent with modulated
reconnection.  \citet{mcintosh2009} provide evidence for 3 to 5
minute periodicities in weak upflows in the transition region in the
quiet sun, which may be linked to similar coronal periodicities.

A very different explanation sometimes put forth for quasi-periodic
behavior in flares and Type III radio `storms' is based on
avalanche or chaos models \citep{eastwood2010,isliker1998}. In a study of two other intervals in this PSP encounter, \citet{Pulupa_2020}found that the power law index of intensity distribution was consistent with \citep{eastwood2010}. They found that the waiting time distribution dependence on frequency differed in the two intervals, discussed possible reasons, and  comparisons to other studies.  Type
III storms with very low amplitude and frequent bursts
($\sim$800/hour) have been reported by \citet{tun2015} 
using high time resolution ground-based instruments. They propose that
continuous reconnection in the corona in concert with density and
temperature inhomogeneities may explain their observations. Studies have
predicted a range of periodicities for nanoflares \citep{viall2011, bradshaw2012, klimchuk2015} 
depending on parameters such
cooling rates. MHD simulations of the time intervals between nanoflare
reconnection. \cite{knizhnik2020} showed a power law distribution,
similar to the conclusion of \cite{eastwood2010} for Type III
storms.~ The time intervals we observed between acceleration events are
very consistent over many day intervals, and thus not explainable by a
process resulting in a power law distribution.

We have reported the first observations of periodic Type III radio
bursts by Parker Solar Probe and their correlation with periodic rapid
heating and cooling in an active region in several EUV channels of the SDO/AIA, and with
variations in sunspot brightness seen in the SDO/HMI. The periods were
$\sim$5 minutes in all wavelengths, and comparable to solar
p-modes. Similar Type III bursts were also observed by WIND. NuSTAR hard
x-rays occurred in association with at least one small microflare in the
active region, but were not directly correlated with the Type III
bursts. Because Type III radio bursts are generated by non-thermal
electrons, this event provides strong evidence for quasi-periodic
small-scale acceleration processes in the corona during quiet times. The
periodic Type III bursts were observed for days, suggesting that these
periodic electron acceleration events may be important for understanding
coronal heating.

\

\begin{acknowledgements}
We acknowledge the NASA Parker Solar Probe Mission,
and the FIELDS team led by S. D. Bale, and the SWEAP team led by J. Kasper
for use of data. The FIELDS experiment on the Parker Solar Probe
spacecraft was designed and developed under NASA contract NNN06AA01C,
and data analysis at UMN and UCB was supported under the same contract.
Work at UMN was also supported by the NASA SolFER Drive Science Center (grant 80NSSC20K0627) and the NASA NuSTAR Guest Observer program (grant 80NSSC18K1744). S.T.B. was supported by NASA Headquarters under the NASA Earth and SpaceScience Fellowship Program Grant 80NSSC18K1201
\end{acknowledgements}

%
%

\bibliographystyle{aa} 
\bibliography{biblio} 

\begin{thebibliography}{66}
\expandafter\ifx\csname natexlab\endcsname\relax\def\natexlab#1{#1}\fi

\bibitem[{Aschwanden(2006)}]{aschwanden2005}
Aschwanden, M. 2006, Physics of the solar corona (Springer Science \& Business
  Media)

\bibitem[{Aschwanden {et~al.}(2016)Aschwanden, Crosby, Dimitropoulou,
  Georgoulis, Hergarten, McAteer, Milovanov, Mineshige, Morales, Nishizuka,
  Pruessner, Sanchez, Sharma, Strugarek, \& Uritsky}]{aschwanden2016}
Aschwanden, M.~J., Crosby, N.~B., Dimitropoulou, M., {et~al.} 2016, Space
  Science Reviews, 198, 47

\bibitem[{Auchère {et~al.}(2016)Auchère, Froment, Bocchialini, Buchlin, \&
  Solomon}]{auchere2016}
Auchère, F., Froment, C., Bocchialini, K., Buchlin, E., \& Solomon, J. 2016,
  The Astrophysical Journal, 825, 110

\bibitem[{Badman {et~al.}(2020)Badman, Bale, Martínez~Oliveros, Panasenco,
  Velli, Stansby, Buitrago-Casas, Réville, Bonnell, Case, Dudok~de Wit, Goetz,
  Harvey, Kasper, Korreck, Larson, Livi, MacDowall, Malaspina, Pulupa, Stevens,
  \& Whittlesey}]{badman2020}
Badman, S.~T., Bale, S.~D., Martínez~Oliveros, J.~C., {et~al.} 2020, The
  Astrophysical Journal Supplement Series, 246, 23

\bibitem[{Bale {et~al.}(2016)Bale, Goetz, Harvey, Turin, Bonnell,
  Dudok de Wit, Ergun, MacDowall, Pulupa, Andre, Bolton, Bougeret, Bowen,
  Burgess, Cattell, Chandran, Chaston, Chen, Choi, Connerney, Cranmer,
  Diaz-Aguado, Donakowski, Drake, Farrell, Fergeau, Fermin, Fischer, Fox,
  Glaser, Goldstein, Gordon, Hanson, Harris, Hayes, Hinze, Hollweg, Horbury,
  Howard, Hoxie, Jannet, Karlsson, Kasper, Kellogg, Kien, Klimchuk,
  Krasnoselskikh, Krucker, Lynch, Maksimovic, Malaspina, Marker, Martin,
  Martinez-Oliveros, McCauley, McComas, McDonald, Meyer-Vernet, Moncuquet,
  Monson, Mozer, Murphy, Odom, Oliverson, Olson, Parker, Pankow, Phan,
  Quataert, Quinn, Ruplin, Salem, Seitz, Sheppard, Siy, Stevens, Summers,
  Szabo, Timofeeva, Vaivads, Velli, Yehle, Werthimer, \& Wygant}]{bale2016}
Bale, S.~D., Goetz, K., Harvey, P.~R., {et~al.} 2016, Space Science Reviews,
  204, 49

\bibitem[{{Battams} {et~al.}(2019){Battams}, {Gallagher}, \&
  {Weigel}}]{battams2019}
{Battams}, K., {Gallagher}, B.~M., \& {Weigel}, R.~S. 2019, \solphys, 294, 11

\bibitem[{Bougeret {et~al.}(1984)Bougeret, Fainberg, \& Stone}]{bougeret1984}
Bougeret, J.-L., Fainberg, J., \& Stone, R. 1984, Astronomy and Astrophysics,
  136, 255

\bibitem[{Bougeret {et~al.}(1995)Bougeret, Kaiser, Kellogg, Manning, Goetz,
  Monson, Monge, Friel, Meetre, Perche, \& {others}}]{bougeret1995}
Bougeret, J.-L., Kaiser, M.~L., Kellogg, P.~J., {et~al.} 1995, Space Science
  Reviews, 71, 231, publisher: Springer

\bibitem[{Bradshaw {et~al.}(2012)Bradshaw, Klimchuk, \& Reep}]{bradshaw2012}
Bradshaw, S.~J., Klimchuk, J.~A., \& Reep, J.~W. 2012, The Astrophysical
  Journal, 758, 53

\bibitem[{De~Moortel {et~al.}(2002)De~Moortel, Ireland, Hood, \&
  Walsh}]{demoortel2002}
De~Moortel, I., Ireland, J., Hood, A.~W., \& Walsh, R.~W. 2002, Astronomy \&
  Astrophysics, 387, L13

\bibitem[{De~Moortel \& Nakariakov(2012)}]{demoortel2012}
De~Moortel, I. \& Nakariakov, V.~M. 2012, Philosophical Transactions of the
  Royal Society A: Mathematical, Physical and Engineering Sciences, 370, 3193

\bibitem[{De~Pontieu {et~al.}(2005)De~Pontieu, Erdélyi, \&
  De~Moortel}]{depontieu2005}
De~Pontieu, B., Erdélyi, R., \& De~Moortel, I. 2005, The Astrophysical Journal
  Letters, 624, L61, publisher: IOP Publishing

\bibitem[{De~Pontieu \& McIntosh(2010)}]{depontieu2010}
De~Pontieu, B. \& McIntosh, S.~W. 2010, The Astrophysical Journal, 722, 1013

\bibitem[{{Dennis} {et~al.}(2017){Dennis}, {Tolbert}, {Inglis}, {Ireland },
  {Wang}, {Holman}, {Hayes}, \& {Gallagher}}]{dennis2017}
{Dennis}, B.~R., {Tolbert}, A.~K., {Inglis}, A., {et~al.} 2017, \apj, 836, 84

\bibitem[{Dominique {et~al.}(2018)Dominique, Zhukov, Dolla, Inglis, \&
  Lapenta}]{dominique2018}
Dominique, M., Zhukov, A.~N., Dolla, L., Inglis, A., \& Lapenta, G. 2018, Solar
  Physics, 293

\bibitem[{Eastwood {et~al.}(2009)Eastwood, Wheatland, Hudson, Krucker, Bale,
  Maksimovic, Goetz, \& Bougeret}]{eastwood2010}
Eastwood, J., Wheatland, M., Hudson, H., {et~al.} 2009, The Astrophysical
  Journal Letters, 708, L95

\bibitem[{Fainberg \& Stone(1970)}]{fainberg1970}
Fainberg, J. \& Stone, R.~G. 1970, Solar Physics, 15, 222

\bibitem[{Fleishman {et~al.}({2008})Fleishman, Bastian, \&
  Gary}]{fleishman2008}
Fleishman, G.~D., Bastian, T.~S., \& Gary, D.~E. {2008}, {ASTROPHYSICAL
  JOURNAL}, {684}, {1433}

\bibitem[{Foullon {et~al.}(2010)Foullon, Fletcher, Hannah, Verwichte, Cecconi,
  Nakariakov, Phillips, \& Tan}]{foullon2010}
Foullon, C., Fletcher, L., Hannah, I.~G., {et~al.} 2010, The Astrophysical
  Journal, 719, 151

\bibitem[{Glesener {et~al.}(2020)Glesener, Krucker, Duncan, Hannah,
  Grefenstette, Chen, Smith, White, \& Hudson}]{glesener2020}
Glesener, L., Krucker, S., Duncan, J., {et~al.} 2020, The Astrophysical
  Journal, 891, L34

\bibitem[{Harra {et~al.}(2020)Harra, Brooks, Bale, Mandrini, Marczynski,
  Sharma, Badman, \& Dominiguez}]{harra2020}
Harra, L., Brooks, D.~H., Bale, S.~D., {et~al.} 2020, Astronomy and
  Astrophysics, submitted

\bibitem[{{Hayes} {et~al.}(2020){Hayes}, {Inglis}, {Christe}, {Dennis}, \&
  {Gallagher}}]{hayes2020}
{Hayes}, L.~A., {Inglis}, A.~R., {Christe}, S., {Dennis}, B., \& {Gallagher},
  P.~T. 2020, \apj, 895, 50

\bibitem[{Hudson(1991)}]{hudson1991}
Hudson, H.~S. 1991, Solar Physics, 133, 357

\bibitem[{{Inglis} {et~al.}(2015){Inglis}, {Ireland}, \&
  {Dominique}}]{inglis2015}
{Inglis}, A.~R., {Ireland}, J., \& {Dominique}, M. 2015, \apj, 798, 108

\bibitem[{Innes {et~al.}(2011)Innes, Cameron, \& Solanki}]{innes2011}
Innes, D.~E., Cameron, R.~H., \& Solanki, S.~K. 2011, Astronomy \&
  Astrophysics, 531, L13

\bibitem[{Ishikawa {et~al.}(2017)Ishikawa, Glesener, Krucker, Christe,
  Buitrago-Casas, Narukage, \& Vievering}]{ishikawa2017}
Ishikawa, S.-n., Glesener, L., Krucker, S., {et~al.} 2017, Nature Astronomy, 1,
  771

\bibitem[{Isliker {et~al.}(1998)Isliker, Vlahos, Benz, \& Raoult}]{isliker1998}
Isliker, H., Vlahos, L., Benz, A., \& Raoult, A. 1998, Astronomy and
  Astrophysics, 336, 371

\bibitem[{James {et~al.}(2017)James, Subramanian, \& Kontar}]{james2017}
James, T., Subramanian, P., \& Kontar, E.~P. 2017, Monthly Notices of the Royal
  Astronomical Society, 471, 89

\bibitem[{Klimchuk(2015)}]{klimchuk2015}
Klimchuk, J.~A. 2015, Philosophical Transactions of the Royal Society A:
  Mathematical, Physical and Engineering Sciences, 373, 20140256

\bibitem[{Knizhnik \& Reep(2020)}]{knizhnik2020}
Knizhnik, K. \& Reep, J. 2020, Solar Physics, 295

\bibitem[{{Krupar} {et~al.}(2020){Krupar}, {Szabo}, {Maksimovic}, {Kruparova},
  {Kontar}, {Balmaceda}, {Bonnin}, {Bale}, {Pulupa}, {Malaspina}, {Bonnell},
  {Harvey}, {Goetz}, {Dudok de Wit}, {MacDowall}, {Kasper}, {Case}, {Korreck},
  {Larson}, {Livi}, {Stevens}, {Whittlesey}, \& {Hegedus}}]{krupar2020}
{Krupar}, V., {Szabo}, A., {Maksimovic}, M., {et~al.} 2020, \apjs, 246, 57

\bibitem[{Kumar {et~al.}(2016)Kumar, Nakariakov, \& Cho}]{kumar2016}
Kumar, P., Nakariakov, V.~M., \& Cho, K.-S. 2016, The Astrophysical Journal,
  822, 7

\bibitem[{Kumar {et~al.}(2017)Kumar, Nakariakov, \& Cho}]{kumar2017}
Kumar, P., Nakariakov, V.~M., \& Cho, K.-S. 2017, The Astrophysical Journal,
  844, 149

\bibitem[{Kupriyanova {et~al.}(2020)Kupriyanova, Kolotkov, Nakariakov, \&
  Kaufman}]{kupriyanova2020}
Kupriyanova, E., Kolotkov, D., Nakariakov, V., \& Kaufman, A. 2020,
  Solar-Terrestrial Physics, 6, 3

\bibitem[{Kupriyanova {et~al.}(2016)Kupriyanova, Kashapova, Reid, \&
  Myagkova}]{kupriyanova2016}
Kupriyanova, E.~G., Kashapova, L.~K., Reid, H. A.~S., \& Myagkova, I.~N. 2016,
  Solar Physics, 291, 3427

\bibitem[{Lemen {et~al.}(2012)Lemen, Title, Akin, Boerner, Chou, Drake, Duncan,
  Edwards, Friedlaender, Heyman, Hurlburt, Katz, Kushner, Levay, Lindgren,
  Mathur, McFeaters, Mitchell, Rehse, Schrijver, Springer, Stern, Tarbell,
  Wuelser, Wolfson, Yanari, Bookbinder, Cheimets, Caldwell, Deluca, Gates,
  Golub, Park, Podgorski, Bush, Scherrer, Gummin, Smith, Auker, Jerram, Pool,
  Soufli, Windt, Beardsley, Clapp, Lang, \& Waltham}]{lemen2012}
Lemen, J.~R., Title, A.~M., Akin, D.~J., {et~al.} 2012, Solar Physics, 275, 17

\bibitem[{Li {et~al.}(2015)Li, Jiang, Yang, Bi, \& Liang}]{li2015}
Li, H.~D., Jiang, Y.~C., Yang, J.~Y., Bi, Y., \& Liang, H.~F. 2015,
  Astrophysics and Space Science, 359

\bibitem[{Liu {et~al.}(2018)Liu, Jin, Downs, Ofman, C.~M.~Cheung, \&
  Nitta}]{liu2018}
Liu, W., Jin, M., Downs, C., {et~al.} 2018, The Astrophysical Journal, 864, L24

\bibitem[{Liu \& Ofman(2014)}]{liu2014}
Liu, W. \& Ofman, L. 2014, Solar Physics, 289, 3233

\bibitem[{Liu {et~al.}(2011)Liu, Zhao, Ofman, Schrijver, Aschwanden,
  De~Pontieu, Tarbell, {et~al.}}]{liu2011}
Liu, W., Zhao, J., Ofman, L., {et~al.} 2011, The Astrophysical Journal Letters,
  736, L13

\bibitem[{McClements \& Fletcher(2009)}]{mcclements2009}
McClements, K.~G. \& Fletcher, L. 2009, The Astrophysical Journal, 693, 1494

\bibitem[{McIntosh \& De~Pontieu(2009)}]{mcintosh2009}
McIntosh, S.~W. \& De~Pontieu, B. 2009, The Astrophysical Journal, 706, L80

\bibitem[{McLaughlin {et~al.}(2018)McLaughlin, Nakariakov, Dominique,
  JelÃ­nek, \& Takasao}]{mclaughlin2018}
McLaughlin, J.~A., Nakariakov, V.~M., Dominique, M., JelÃ­nek, P., \&
  Takasao, S. 2018, Space Science Reviews, 214

\bibitem[{McLaughlin {et~al.}(2012)McLaughlin, Verth, Fedun, \&
  ErdÃ©lyi}]{mclaughlin2012}
McLaughlin, J.~A., Verth, G., Fedun, V., \& ErdÃ©lyi, R. 2012, The
  Astrophysical Journal, 749, 30

\bibitem[{Morioka {et~al.}(2007)Morioka, Miyoshi, Masuda, Tsuchiya, Misawa,
  Matsumoto, Hashimoto, \& Oya}]{morioka2007}
Morioka, A., Miyoshi, Y., Masuda, S., {et~al.} 2007, The Astrophysical Journal,
  657, 567

\bibitem[{Murray {et~al.}(2009)Murray, van Driel-Gesztelyi, \&
  Baker}]{murray2009}
Murray, M.~J., van Driel-Gesztelyi, L., \& Baker, D. 2009, Astronomy \&
  Astrophysics, 494, 329

\bibitem[{Nakariakov {et~al.}({2006})Nakariakov, Foullon, Verwichte, \&
  Young}]{nakariakov2006}
Nakariakov, V.~M., Foullon, C., Verwichte, E., \& Young, N.~P. {2006},
  {ASTRONOMY \& ASTROPHYSICS}, {452}, {343}

\bibitem[{Nakariakov \& Melnikov(2009)}]{nakariakov2009}
Nakariakov, V.~M. \& Melnikov, V.~F. 2009, Space Science Reviews, 149, 119

\bibitem[{Nakariakov \& Verwichte(2005)}]{nakariakov2005}
Nakariakov, V.~M. \& Verwichte, E. 2005, Living Reviews in Solar Physics, 2

\bibitem[{Ofman \& Thompson(2002)}]{ofman2002}
Ofman, L. \& Thompson, B.~J. 2002, The Astrophysical Journal, 574, 440

\bibitem[{Parks \& Winckler(1969)}]{parks1969}
Parks, G. \& Winckler, J. 1969, The Astrophysical Journal, 155, L117

\bibitem[{Pulupa {et~al.}(2020)Pulupa, Bale, Badman, Bonnell, Case, de~Wit,
  Goetz, Harvey, Hegedus, Kasper, Korreck, Krasnoselskikh, Larson, Lecacheux,
  Livi, MacDowall, Maksimovic, Malaspina, Oliveros, Meyer-Vernet, Moncuquet,
  Stevens, \& Whittlesey}]{Pulupa_2020}
Pulupa, M., Bale, S.~D., Badman, S.~T., {et~al.} 2020, The Astrophysical
  Journal Supplement Series, 246, 49

\bibitem[{Pulupa {et~al.}(2017)Pulupa, Bale, Bonnell, Bowen, Carruth, Goetz,
  Gordon, Harvey, Maksimovic, Martínez-Oliveros, Moncuquet, Saint-Hilaire,
  Seitz, \& Sundkvist}]{pulupa2017}
Pulupa, M., Bale, S.~D., Bonnell, J.~W., {et~al.} 2017, Journal of Geophysical
  Research: Space Physics, 122, 2836

\bibitem[{Roberts(2000)}]{roberts2000}
Roberts, B. 2000, Solar Physics, 193, 139

\bibitem[{Roberts {et~al.}(1984)Roberts, Edwin, \& Benz}]{roberts1984}
Roberts, B., Edwin, P., \& Benz, A. 1984, The Astrophysical Journal, 279, 857

\bibitem[{Schou {et~al.}(2012)Schou, Scherrer, Bush, Wachter, Couvidat,
  Rabello-Soares, Bogart, Hoeksema, Liu, Duvall, Akin, Allard, Miles, Rairden,
  Shine, Tarbell, Title, Wolfson, Elmore, Norton, \& Tomczyk}]{schou2012}
Schou, J., Scherrer, P.~H., Bush, R.~I., {et~al.} 2012, Solar Physics, 275, 229

\bibitem[{Sych {et~al.}(2012)Sych, Zaqarashvili, Nakariakov, Anfinogentov,
  Shibasaki, \& Yan}]{sych2012}
Sych, R., Zaqarashvili, T.~V., Nakariakov, V.~M., {et~al.} 2012, Astronomy \&
  Astrophysics, 539, A23

\bibitem[{Sych {et~al.}(2020)Sych, Zhugzhda, \& Yan}]{sych2020}
Sych, R., Zhugzhda, Y., \& Yan, X. 2020, The Astrophysical Journal, 888, 84

\bibitem[{Thurgood {et~al.}(2017)Thurgood, Pontin, \&
  McLaughlin}]{thurgood2017}
Thurgood, J.~O., Pontin, D.~I., \& McLaughlin, J.~A. 2017, The Astrophysical
  Journal, 844, 2

\bibitem[{Tun~Beltran {et~al.}(2015)Tun~Beltran, Cutchin, \& White}]{tun2015}
Tun~Beltran, S.~D., Cutchin, S., \& White, S. 2015, Solar Physics, 290, 2423

\bibitem[{Van Doorsselaere {et~al.}(2016)Van Doorsselaere, Kupriyanova, \&
  Yuan}]{vandoorsselaere2016}
Van Doorsselaere, T., Kupriyanova, E.~G., \& Yuan, D. 2016, Solar Physics,
  291, 3143

\bibitem[{Veronig {et~al.}(2000)Veronig, Messerotti, \&
  Hanslmeier}]{veronig2000}
Veronig, A., Messerotti, M., \& Hanslmeier, A. 2000, Astronomy \& Astrophysics,
  357, 337, \_eprint: nlin/0207021

\bibitem[{Veronig {et~al.}(2011)Veronig, Gömöry, Kienreich, Muhr, Vršnak,
  Temmer, \& Warren}]{veronig2011}
Veronig, A.~M., Gömöry, P., Kienreich, I.~W., {et~al.} 2011, The
  Astrophysical Journal, 743, L10

\bibitem[{Viall \& Klimchuk(2011)}]{viall2011}
Viall, N.~M. \& Klimchuk, J.~A. 2011, The Astrophysical Journal, 738, 24

\bibitem[{Zhao {et~al.}(2016)Zhao, Felipe, Chen, \& Khomenko}]{zhao2016}
Zhao, J., Felipe, T., Chen, R., \& Khomenko, E. 2016, The Astrophysical Journal
  Letters, 830, L17

\bibitem[{Zharkova {et~al.}(2011)Zharkova, Arzner, Benz, Browning, Dauphin,
  Emslie, Fletcher, Kontar, Mann, Onofri, Petrosian, Turkmani, Vilmer, \&
  Vlahos}]{zharkova2011}
Zharkova, V.~V., Arzner, K., Benz, A.~O., {et~al.} 2011, Space Science Reviews,
  159, 357

\end{thebibliography}

\end{document}